\documentclass{article}

\usepackage{amsmath}
\usepackage{amsfonts}
\usepackage{amsthm}
\usepackage{amsopn}
\usepackage{verbatim}

\newcommand{\lb}{\left (}
\newcommand{\rb}{\right )}
\newcommand{\B}[1]{\lb #1 \rb}
\newcommand{\lsb}{\left [}
\newcommand{\rsb}{\right ]}

\newcommand{\be}{\begin{eqnarray}}
\newcommand{\ee}{\end{eqnarray}}
\newcommand{\bc}{}

\newcommand{\me}[0]{\mathcal{E}}

\newcommand{\YAMLSectionize}[1]{}

\def\classlim{q \rightarrow 1}

\newcommand{\Cartan}[1]{\lsb q^{2\phi_i H_i} \rsb}

\hoffset= - 1.0in         

\title{{\bf A note on quantum groups and integrable systems} \vspace{.2cm}}
\author{{\bf A.Popolitov}}

\begin{document}
\date{}
 \maketitle

\vspace{-5.0cm}

\begin{center}
\hfill IITP/TH-19/15\\
\end{center}

\vspace{3.5cm}

\begin{center}

{\small {\it ITEP, Moscow 117218, Russia}}\\
{\small {\it Institute for Information Transmission Problems, Moscow 127994, Russia}}\\
{\small {\it Korteweg-de Vries Institute for Mathematics, University of
Amsterdam, \\P.O. Box 94248, 1090 GE Amsterdam, The Netherlands}}
\end{center}

\centerline{ABSTRACT}

\bigskip

{\footnotesize
  Free-field formalism for quantum groups \cite{MV} provides a special choice
  of coordinates on a quantum group. In these coordinates the construction of
  associated integrable system \cite{Mars1} is especially simple.
  This choice also fits into general framework of cluster varieties \cite{FG1}
  -- natural changes of coordinates are cluster mutations.
}


\bigskip

\bigskip

Very often we describe physical systems using perturbation theory.
But to develop a perturbation theory, we need some starting point,
the model that can be solved exactly -- the integrable system.
When we encounter new physical system, we need to choose this starting point
-- and the wider choice we have, the better.
Hence, search for integrable systems is very important.

Usually, a system is integrable
because it has some symmetry (see e.g. \cite{M1}, \cite{M2}, \cite{AASZ} and \cite{LOZ1}).
If there is a continuum of symmetries, they form a Lie group.
If the system is quantum, then symmetries can form quantum Lie group.

But connection between Lie groups and integrable systems is even more strong.
We can actually use a classical Lie group to construct a bunch of
integrable systems \cite{FM}. The phase spaces of these integrable systems
are Poisson submanifolds (leafs) of dimension $2 n$, where $n$ is the rank of the group.
The integrals of motion are Ad-invariant functions on the group, and there are exactly $n$ of them.

\bigskip

Now one may wonder: \textit{does this construction of integrable systems extend to quantum groups?}
This note outlines, how to do this, for the case of quantum group $SL_q(N)$.

The question is harder than it seems.
This is because quantum group has a rather technical and indirect definition.
One starts with $q$-deforming the Lie algebra commutational relations
in the so-called Chevalley basis.
Then one considers all possible functions of these elements.
In this space of all functions, one is interested only in group-like elements
-- functions $g$, that satisfy equation
\be \label{eq:group-like}
\Delta(g) = g \otimes g,
\ee
where $\Delta$ is the comultiplication -- another important component of a
quantum group definition. Solutions of this equation form a group and
as $q \rightarrow 1$ this set of solutions coincides with classical group $SL(N)$.
Thus, it is this set of group-like elements, that is the proper $q \neq 1$ deformation of a Lie group $SL(N)$.

The story is even more tricky than this.
If one considers Taylor coefficients to be commutative (e.g. complex numbers),
then equation \eqref{eq:group-like} does not have interesting solutions.
So, one needs to introduce extra $N^2 - 1$ non-commutative parameters,
which become commutative in the classical limit.
The question is how to introduce these non-commutative variables in such a
way, that the result has some physically interesting properties.

\bigskip

A clever approach to this problem, which is simultaneously
a better, more direct way to work with group-like variety,
is to propose some ansatz for group-like element in terms of new non-commutative parameters \cite{MV}.
Equation \eqref{eq:group-like} then would imply some commutational
relations for these parameters.
There are two crucial features of the Morozov-Vinet ansatz.
First, while it uses $N^2 - 1$ parameters, it only uses \textit{simple} roots of
the quantum Lie algebra (there are $2N-2$ of them), but each one multiple times.
Second, $q \rightarrow 1$ limit of this ansatz looks like exponential map
from Lie algebra to its Lie group. As a result, the required commutational
relations have very simple form

\be \label{eq:mv-comm-rels}
& \psi_i \psi_j = q^{-C_{[i][j]}} \psi_j \psi_i, \ \ 
\chi_i \chi_j = q^{-C_{[i][j]}} \chi_j \chi_i, \text{ for } i < j, \\ \nonumber
& q^{\phi_i} \psi_j = q^{C[i][j]} \psi_j q^{\phi_i}, \ \
q^{\phi_i} \chi_j = q^{C[i][j]} \chi_j q^{\phi_i}, \\ \nonumber
& \chi_i \psi_j = \chi_j \psi_i, \text{ for all } i,\ j,
\ee
where $\frac{N(N-1)}{2}$ parameters $\psi_i$ are coupled to negative roots,
$\frac{N(N-1)}{2}$ parameters $\chi_i$ -- to positive roots,
and $N-1$ parameters $\phi_i$ -- to Cartan elements, $C_{ij}$
is the $SL_q(N)$ Cartan matrix.
The bracket map $[i]$ gives the number of the simple root,
associated with the $i$-th variable.
For example, for $SL_q(4)$ we have: $[1] = 1$, $[2] = 2$, $[3] = 1$, $[4] = 1$, $[5] = 2$, $[6] = 1$.
These commutational relations are exponents of Heisenberg-like commutational relations,
which motivates the name ``free-field formalism'' -- and makes this approach so attractive.

\bigskip

Even though Morozov-Vinet ansatz already leads to simple commutational relations,
it is still not clear how to reduce system to $2 n$-dimensional subspace.
But at $q = 1$ this ansatz is very similar
to explicit parametrization of classical Lie group element, used in \cite{FM} and \cite{FG1,FG2}
-- with a slight difference -- Cartan elements are interspersed with simple roots.
For example, in case of $SL(3)$ parametrization of \cite{FM} looks like:

\be
g = H_1(x^1_0) E_1 H_1(x^1_1) F_1 H_1(x^1_2) \Bigg( H_2(x^2_0) E_2 H_2(x^2_1) F_2 H_2(x^2_2) \Bigg) E_1 H_1(x^1_3) F_1 H_1(x^1_4),
\ee
where $E_i = \exp(e_i), F_i = \exp(f_i)$ -- exponents of positive and negative simple roots
and $H_i(x) = x^{h^i}$ -- exponents of Cartan elements. Structurally this ansatz looks
like a word of numbers $1\bar{1}2\bar{2}1\bar{1}$ -- if $i$ stands for $E_i$ and $\bar{i}$ stands for $F_i$,
with appropriate $H(x)$ inserted between every two numbers (for details, see \cite{FG2}).

We can use this ansatz as a hint, how to improve Morozov-Vinet ansatz for $q \neq 1$
to make integrable system construction more explicit.
The only adjustments, we need to do, is to change exponential functions by appropriate $q$-exponential functions
and (this is a technical detail, related to exact formula for comultiplication),
unlike \cite{FG2}, but more like \cite{MV}, consider $H(x) = x^{h_i},$ where
$h_i = 1/2 C_{ij} h^j$.
Then same group variety equation \eqref{eq:group-like} will lead to even simpler commutational
relations

\be \label{eq:comm-rels}
& x^1_0 x^1_1 = q^{-2} x^1_0 x^1_1, \ \ x^1_1 x^1_2 = q^{2} x^1_2 x^1_1, \ \ x^1_2 x^1_3 = q^{-2} x^1_3 x^1_2,
\ \ x^1_3 x^1_4 = q^{2} x^1_4 x^1_3, \\ \nonumber
& x^2_0 x^2_1 = q^{-2} x^2_0 x^2_1, \ \ x^2_1 x^2_2 = q^{2} x^2_2 x^2_1,
\ee
and all the other commutators are zero. This can be conveniently encoded in directed graphs

\be \label{eq:cluster-picture} &
\begin{picture}(100,20)(0,-10)
  \put(0,0){\circle*{4}}
  \qbezier(0,0)(10,0)(20,0) \put(12,0){\vector(1,0){0}}
  \put(20,0){\circle*{4}}
  \put(20,0){\qbezier(0,0)(10,0)(20,0)} \put(28,0){\vector(-1,0){0}}
  \put(40,0){\circle*{4}}
  \put(40,0){\qbezier(0,0)(10,0)(20,0)} \put(52,0){\vector(1,0){0}}
  \put(60,0){\circle*{4}}
  \put(60,0){\qbezier(0,0)(10,0)(20,0)} \put(68,0){\vector(-1,0){0}}
  \put(80,0){\circle*{4}}
  \put(-5,-10){$x^1_0$}
  \put(15,-10){$x^1_1$}
  \put(35,-10){$x^1_2$}
  \put(55,-10){$x^1_3$}
  \put(75,-10){$x^1_4$} 
\end{picture} \\ \nonumber
&
\begin{picture}(100,20)(0,-10)
  \put(0,0){\circle*{4}}
  \qbezier(0,0)(10,0)(20,0) \put(12,0){\vector(1,0){0}}
  \put(20,0){\circle*{4}}
  \put(20,0){\qbezier(0,0)(10,0)(20,0)} \put(28,0){\vector(-1,0){0}}
  \put(40,0){\circle*{4}}
  \put(-5,-10){$x^2_0$}
  \put(15,-10){$x^2_1$}
  \put(35,-10){$x^2_2$}
\end{picture}
\ee
which generalizes to $SL_q(N)$ case in an obvious way
(for $SL_q(4)$ there will be 3 graphs with 3, 5 and 7 vertices, respectively).
 We see that variables split
into $N-1$ groups, which do not interact with each other.
It is straightforward to see (taking into account difference between $h_i$ and $h^i$) that
these commutational relations reproduce Poisson bracket of \cite{FG2}
 in $\classlim$ limit.

\bigskip

In these variables it is easy to see how to do reduction to $2n$-leaf.
It is clear, that if we consider only functions that depend
(in $SL_q(3)$ case) on $x^1_0$, $x^1_1$, $x^2_0$ and $x^2_1$
(i.e. on first two coordinates coupled to each simple root, for general $N$),
the space of such functions is closed under commutational relations \eqref{eq:comm-rels}.
This means, that it defines a $2n$-dimensional submanifold in quantum group $SL_q(N)$.
Points of this manifold can be explicitly parametrized as

\begin{equation}
g_{2n} = \prod_{i = 1}^{N-1} \B{x^i_0}^{h_i} \me_q \B{e_i} \B{x^i_1}^{h_i} \me_{1/q} \B{f_i},
\end{equation}
where $\me_q(x)$ is the $q$-exponential.
Taking the limit $\classlim$ in this expression, we obtain
formula for $2n$ symplectic manifold of a classical Lie group from \cite{Mars1},
on which a relativistic Toda chain lives.
Thus, it is natural to think, that on $g_{2n}$ is the phase space
of $q$-deformed relativistic Toda chain.

\bigskip

Finally, looking at commutational relations \eqref{eq:comm-rels} we see, that the
structure of cluster variety, which is present at $q = 1$ survives quantization and
is promoted to structure of quantum cluster variety \cite{FG1}.
Actually, if, for example, we change the order of first positive and first negative root in the quantum ansatz,
so that it becomes

\be
g = H_1(y^1_0) F_1 H_1(y^1_1) E_1 H_1(y^1_2) \lb H_2(x^2_0) E_2 H_2(x^2_1) F_2 H_2(x^2_2) \rb E_1 H_1(x^1_3) F_1 H_1(x^1_4),
\ee
where $H_i(x) = x^{h_i}$, $E_i = \me_q(e_i)$ and $F_i = \me_{1/q}(f_i)$,
we would get that this new parametrization is related to the old one by simple non-commutative
change of variables

\be\label{eq:mutation}
y^1_0 = x^1_0 (1 + q x^1_1), \ \ y^1_2 = x^1_2 (1 + q x^1_1), \ \ y^1_1 = 1 / x^1_1
\ee

Changes of variables of this form are precisely cluster mutations and are, in fact, also
encoded in the picture \eqref{eq:cluster-picture}.

\bigskip

To summarize, even though there are a lot of peculiarities and details that need to
be kept in mind when doing practical calculations, the general picture is clear and simple.
Construction of integrable systems of \cite{FM} generalizes to $q \neq 1$ case
almost literally. There it no longer looks \textit{ad hoc} -- all the choices are
fixed by group variety equation \eqref{eq:group-like}, an essential component of
the quantum group. Furthermore, the rich structure of cluster variety is fully
present in the quantum case -- commutational relations and natural changes of coordinates
(mutations)
are encoded in cluster quivers like \eqref{eq:cluster-picture}. Even though we performed
our calculations only for $SL_q(N)$ group we hope everything is true for other Lie groups
-- but this is subject of future research.

\section*{Acknowledgments}

Author wants to thank A.Mironov and A.Morozov for many stimulating discussions.
This work is partly supported by grant for support of scientific schools
NSh-1500.2014.2, RFBR grants 15-31-20832-mol-a-ved
and 14-02-00627, by 15-52-50041-YaF.

\appendix

\end{document}